\documentclass[a4paper,12pt]{article}

\usepackage{jheppub}

\usepackage{hyperref}

\def\be{\begin{equation}}
\def\ee{\end{equation}}
\def\ba{\begin{eqnarray}}
\def\ea{\end{eqnarray}}
\def\bi{\begin{itemize}}
\def\ei{\end{itemize}}

\def\O{\Omega}
\def\xh{\hat{x}}

\def\e{\varepsilon}
\def\qh{\hat{q}}

\def\zb{\bar{z}}
\def\w{\omega}

\def\scri{\mathcal{I}}
\def\lamtwo{\overset{(2)}{\lambda}}
\def\lamone{\overset{(1)}{\lambda}}
\def\lamzero{\overset{(0)}{\lambda}}
\def\lamln{\overset{\scriptscriptstyle (\log r/r)}{\lambda}}

\def\lam{\lambda}
\def\A{\mathcal{A}}

\def\J{\mathcal{J}}
\def\j{\mathcal{J}}
\def\F{\mathcal{F}}
\def\I{\mathcal{I}}
\def\Fo{\overset{ (0)}{\F}}
\def\Fone{\overset{ (-1)}{\F}}
\def\Ftwo{\overset{ (-2)}{\F}}
\def\Fthree{\overset{(-3)}{\F}}
\def\jtwo{\overset{(-2)}{\j}}

\def\rhodiv{\rho_{\text{div}}}
\def\rhofin{\rho_{\text{finite}}}
\def\Qt{\tilde{Q}}

\def\D{\mathcal{D}}
\def\phitwo{\overset{(-2)}{\varphi}}
\def\Atwo{\overset{(-2)}{\A}}
\def\rhot{\tilde{\rho}}
\def\et{\tilde{\varepsilon}}
\def\mut{\tilde{\mu}}
\def\rhotdiv{\tilde{\rho}_{\text{div}}}
\def\rhotfin{\tilde{\rho}_{\text{finite}}}
\def\Q{\mathcal{Q}}
\def\Qs{\Q^{\rm soft}}
\def\Qh{\Q^{\rm hard}}

\title{Subleading soft photons and large gauge transformations}
\author[a]{Miguel Campiglia}
\author[b]{Alok Laddha}
\affiliation[a]{Instituto de F\'{\i}sica, Facultad de Ciencias, 
Igu\'a 4225, Montevideo, Uruguay.}
\affiliation[b]{Chennai Mathematical Institute, Siruseri 603103, India}

\emailAdd{campi@fisica.edu.uy}
\emailAdd{aladdha@cmi.ac.in}

\abstract{
Lysov, Pasterski and Strominger  have shown how Low's subleading soft photon theorem can be understood as Ward identities of  new  symmetries of  massless QED.
In this paper we offer a different perspective 
and show that there exists a class of large $U(1)$ gauge transformations such that  (i) the associated  (electric and magnetic)  charges can be computed from first principles, (ii) their  Ward identities are equivalent to Low's theorem.
Our framework  paves the way to analyze the sub-subleading theorem in gravity in terms of Ward identities associated to large diffeomorphisms. 
}

\begin{document}
\maketitle

\section{Introduction}

The analysis of  asymptotic symmetries in gauge theories and gravity has seen a resurgence in the last few years due to the seminal work of Strominger and collaborators \cite{strom0}.   
In particular, it was shown in \cite{stromqed}
that the classic Weinberg's soft photon theorem  can be understood as a Ward identity associated to an infinite dimensional symmetry group of QED. This group is obtained by considering 
large  gauge transformations at null infinity, and implies an infinite number of conservation laws in the scattering processes. The analysis in \cite{stromqed}, originally in the context of massless particles, was  later extended  to  massive particles \cite{cl3,strommass}  thereby strengthening the overall picture. 
However, the factorization constraints on QED  extend beyond  leading order. As shown by Low \cite{low} the factorization of scattering amplitudes applies also to the next order in the photon energy. 
The theorem takes the form \cite{stromlow}:
\begin{equation}
\lim_{\omega\rightarrow\ 0}(1 + \omega\partial_{\omega}){\cal M}_{n+1}(k_{1},\dots,k_{n};\w \qh) \ =\ S^{(1)}{\cal M}_{n}(k_{1},\dots,k_{n}) , \label{lowthm}
\end{equation}
where $S^{(1)}$ is a sum of differential operators acting on the external momenta $k_i$. While $S^{(1)}$ is expected to be sensitive to loop corrections \cite{stromqed} our focus will be on the  theorem at tree level.

A natural question, first investigated by Lysov, Pasterski and Strominger  \cite{stromlow} is whether  Low's theorem  can  also be understood as  Ward identities. In \cite{stromlow} the authors showed that the theorem is equivalent to Ward identities of infinitely many charges that are parametrized by vector fields on the sphere.  They interpreted the charges as local  generalizations of electric and magnetic dipole moments.  In this paper we offer an alternative perspective on this charges and show that in fact  they are associated to certain large $U(1)$ gauge transformations.

This work is a precursor to \cite{short} where we apply the same conceptual ideas to the case of gravity and show that there exists a new class of symmetries whose Ward identities are equivalent to the sub-subleading soft graviton theorem \cite{cs}.\\
\subsection{Summary of results}
In this section we summarize the key ideas and  results of the paper. We consider massless scalar QED and work in harmonic gauge. This is rather convenient since the soft theorems are usually formulated in this gauge. Global symmetries can then arise  from residual, large gauge transformations  which are parametrized by solutions of
the wave equation
\be\label{boxlam}
\square\lambda\ =\ 0 .
\ee
In retarded $(u,r,\hat{x})$ coordinates, one can solve this equation in an $r \to \infty$ expansion once the asymptotic behavior of $\lambda$ is specified. Typically, the leading component in this expansion provides  ``free data" in terms of which the solution is determined.

For a given large gauge parameter
$\lambda$, one can associate charges of electric and magnetic type according to:
\be
Q_\lambda = \int_\Sigma d^3 V  \partial_a (\lambda E^a), \quad \quad  \Qt_\lambda = \int_\Sigma d^3 V  \partial_a (\lambda B^a), \label{elmagch}
\ee
where $E^a$ and $B^a$ are the electric and magnetic fields with respect to the hypersurface $\Sigma$. 
These charges can be computed on any spatial slice $\Sigma$.  By pushing $\Sigma$ to null infinity the charges become functions on the radiative phase space of the theory \cite{as,aabook} and whence  especially convenient for studying conserved quantities in scattering processes. 

It is the electric-type charge $Q_\lambda$  that has been mostly used in the studies relating   soft theorems with  Ward identities. A notable exception is \cite{strommag}, where  the magnetic-type charge $\tilde{Q}_\lambda$  is used to include the effects of magnetic monopoles.
Here we will show that  $\tilde{Q}_\lambda$   plays a key role  already in the ordinary case where no magnetic monopoles are present.
This can already  be seen in the large gauge transformations considered in \cite{stromqed}. There, in order to establish the equivalence of (electric) Ward identities with Weinberg's soft photon theorem, certain condition is imposed on the fields which effectively sets to zero the magnetic-type charges. In section \ref{secO1} we reinterpret Weinberg's soft theorem as a Ward identity of both electric and magnetic-type charges with no such condition on the fields. 
 



However our main interest in this paper is  to relate Low's subleading soft photon theorem with large gauge transformation. A first guess based on simple Fourier space reasoning suggests one should look at large gauge parameters whose $O(r^{0})$ component is linear in $u$. 
It turns out that in order for this to be compatible with Eq.(\ref{boxlam}),  the gauge parameter must have an $O(r)$ piece. We show that such solutions exist (at least asymptotically) and compute the corresponding charges at null infinity. These are divergent, but by projecting out a soft photon contribution the charges are rendered finite. 
We then show that these finite charges are  equivalent to the charges obtained in \cite{stromlow}. This in turn establishes the equivalence of the (electric and magnetic) Ward identities  with Low's subleading soft photon theorem.


At this point, a natural question arises. Can one keep going and find more Ward identities? Are there $O(r^2)$ large gauge parameters yielding novel relations for  sub-subleading photons?
In section \ref{secOr2} we argue in the negative and provide evidence that the  $O(1)$ and $O(r)$ gauge parameters exhaust all possible large gauge symmetries.

\section{Preliminaries} \label{secprel}
We consider a massless charged scalar field $\varphi$ coupled to the Maxwell field $\A_\mu$ satisfying the the field equations
 \ba
\nabla^\rho \F_{\mu \rho} & = & \j_\mu ,  \label{eom} \\
\quad \quad  \D_\mu \D^\mu \varphi & = &0, \label{eom2}
\ea
where $\nabla_\mu$ is the spacetime covariant derivative, $\F_{\mu \nu}= \partial_\mu \A_\nu - \partial_\nu \A_\mu$ the field strength, 
\be
\j_\mu= i e \varphi (\D_\mu \varphi)^* + c.c.
\ee
 the charge current and $\D_\mu$ the gauge covariant derivative, $\D_\mu \varphi = \partial_\mu \varphi - i e \A_\mu \varphi$. Local $U(1)$ gauge transformations are parametrized by a scalar $\lam$ and act as
\be
\delta_{\lam} \A_\mu  =  \partial_\mu \lam, \quad \delta_{\lam} \varphi =  i e \lam \varphi \label{gge}.
\ee
In particular, under a gauge transformation the covariant derivative transforms by $\delta_{\lam} \D_\mu \varphi = i e \lambda \D_\mu \varphi$. As in \cite{cl3} we work in harmonic gauge $\nabla^\mu \A_\mu=0$ so that gauge parameters to be considered will satisfy the wave equation (\ref{boxlam}).

As in our previous studies we follow a covariant phase space approach to compute the charges \cite{abr,leewald} . The symplectic form for the system is given by an integral over a Cauchy slice $\Sigma$ (which we  eventually take its limit to null infinity):
\be
\O(\delta,\delta')= \int_\Sigma dS_\mu ( \delta \theta^\mu(\delta') - \delta \leftrightarrow \delta'), 
\ee
where $\theta^\mu$ is the symplectic potential density
\be
\theta^\mu(\delta)= \sqrt{g}(F^{\mu \nu} \delta \A_\nu + (\D^\mu \varphi)^* \delta \varphi + c.c.) \label{deftheta}.
\ee
The generator $Q_{\lam}$ of the gauge transformation (\ref{gge}) is  defined by the condition
\be
\delta Q_\lambda = \O(\delta_\lambda,\delta). \label{delQ}
\ee
Using the field equations (\ref{eom}) one can verify that 
\be
Q_\lambda := - \int_\Sigma dS_\mu \partial_\nu( \sqrt{g} \lambda \F^{\mu \nu}), \label{Qlam}
\ee
satisfies the defining condition (\ref{delQ}). For $\lambda= $ constant this gives the total electric charge of the system. We will also be interested in charges that are the dual,  `magnetic' version of (\ref{Qlam}),
\be
\Qt_{\lambda} := \int_\Sigma dS_\mu \partial_\nu(   \eta^{\mu \nu \alpha \beta}\lambda \F_{\alpha \beta}), \label{Qlamt}
\ee  
where $\eta^{\mu \nu \alpha \beta}$ is the totally antisymmetric symbol. For $\lambda= $ constant this gives the total magnetic charge of the system. Note that, unlike \cite{strommag}, we are not considering magnetic monopoles and the total magnetic charge will always be zero. However, there can be nonzero local magnetic flux and (\ref{Qlamt}) can be nontrivial for non-constant $\lambda$. The charges (\ref{Qlam}) and (\ref{Qlamt}) are the `electric-type' and `magnetic-type' charges of Eq. (\ref{elmagch}).

In this paper we will be evaluating various charges of the type (\ref{Qlam}), (\ref{Qlamt}) in the limit where the Cauchy slice $\Sigma$ approaches null infinity. The relevance of these charges will be in their relation to soft photons theorems. An interesting question, which is outside of the scope of the present work, is how the field equations (with appropriate boundary conditions)  imply the conservations of these charges, as for instance discussed in \cite{stromqed,strommass} for the  $\lambda=O(1)$ case. \\

\noindent For concreteness we focus on future null infinity. We  work in retarded coordinates $(u,r,x^A)$ in terms of which the Minkowski line element reads
\be
ds^2= - du^2 - 2 du dr +r^2 q_{AB} dx^A d x^B, \label{minkmetric}
\ee
where $A=1,2$ are sphere indices and $q_{AB}$ the unit sphere metric. Charges will be computed by choosing $\Sigma$ to be a $t = u + r =$ constant slice  and taking $t$ to infinity, with $u$ and $\xh$ constant. For  a $t=$constant slice, the quantity being integrated in (\ref{Qlam}) is:
\ba
\rho_\lam & := & - \partial_\mu( \sqrt{g} \lambda( \F^{r \mu}+ \F^{u \mu})) \\
&=& \sqrt{q} [\partial_r( r^2 \lambda \F_{u r})- \partial_u (r^2 \lambda \F_{u r}) )]+ r^2 \partial_A(\sqrt{q} \lambda \F_{u}^{\phantom{u}A}) \label{rholam} ,
\ea
where we used that  $\sqrt{g}= r^2 \sqrt{q}$ and that the inverse Minkowski metric in retarded coordinates has nonzero components $g^{rr}=1$, $g^{ur}=g^{ru}=-1$, $g^{AB}= r^{-2}q^{AB}$.
Only the first two terms in  (\ref{rholam}) contribute to the charge since the last one vanishes after integration on the sphere.
For $\Qt_{\lambda}$ (\ref{Qlamt}) the quantity being integrated is:
\ba
\rhot_{\lambda} &:= & \partial_\mu(   (\eta^{r \mu \alpha \beta}+ \eta^{u \mu \alpha \beta}) \lambda \F_{\alpha \beta}) \label{rholamt0} \\
&= & \partial_r( \eta^{AB} \lambda \F_{AB} ) - \partial_u( \eta^{AB} \lambda \F_{AB}) + 2 \partial_A(  \eta^{AB} \lambda (\F_{Br}-\F_{Bu})), \label{rholamt}
\ea
where we are using the convention that $\eta^{u r A B}= \eta^{AB}$ with $\eta^{AB}$ the antisymmetric symbol on the sphere. Again, only the first two terms in (\ref{rholamt})  contribute to charge,  the last one being a total sphere divergence.\\

\noindent We conclude the preliminaries by discussing the $r \to \infty$ fall-offs of the fields. We assume the standard power series expansion  (see for instance \cite{stromqed}):
\be \label{fallF}
\begin{array}{l}
\F_{AB}= F_{AB} + O(r^{-1}), \quad  \F_{ru} =  \Ftwo_{ru}/r^{2}+ \Fthree_{ru}/r^{3}+O(r^{-4}) \\
\F_{A r} =  \Ftwo_{A r}/r^{2}+ O(r^{-3}) , \quad  \F_{A u} = \Fo_{A u} + O(r^{-1}).
\end{array} 
\ee
Here and in the following, it is understood that the coefficients of the $1/r$ expansion are functions of $u$ and $\xh$. Superscripts indicate the corresponding power of $r$. To simplify later expressions, some of the leading coefficients are written with no superscripts and different font style, e.g. $F_{AB} \equiv \Fo_{AB}$. Fall-offs for $\A_\mu$ compatible with (\ref{fallF}) and the gauge condition $\nabla^\mu \A_\mu=0$ are \cite{cl3}:
\be
\A_A = A_A + O(r^{-1}), \quad \A_u = O(r^{-1}), \quad \A_r = O(r^{-2}),
\ee
where $A_A \equiv \overset{(0)}{\A_A}$ plays the role of free data for the Maxwell field. For the scalar field we have
\be
\varphi = \phi/r + O(r^{-2})
\ee
with $\phi  \equiv \overset{(-1)}{\varphi}$ playing the role of free data. These in turn imply the following fall-offs on the charge current:
\be
\j_u =  j_u/r^2 +  O(r^{-3}), \quad \j_A = j_A/r^2 +  O(r^{-3}) , \quad \j_r= j_r/r^4+O(r^{-5}),
\ee
with
\ba
j_u & = &  i e \, \phi \, \partial_u \phi^* + c.c. \\
j_A  & = & i e \, \phi \, (\partial_A +i e A_A) \phi^* + c.c \label{jA} \\
j_r  & = & i e  \, \phi^* \phitwo + c.c. -2 e^2 |\phi|^2 \Atwo_r. \label{jr}
\ea

\section{$O(1)$ large gauge transformation and associated charges} \label{secO1}
In this section we review the charges associated to large gauge transformations with asymptotic behaviour \cite{stromqed},
\be
\lambda(u,r,\xh) = \e(\xh) + O(r^{-\epsilon}). \label{lamlead}
\ee
In appendix \ref{largeggeapp} we show that conditions (\ref{boxlam}), (\ref{lamlead}) can be satisfied to $O(r^{-1})$ and determine the asymptotic form of the subleading term in (\ref{lamlead}) (which turns out to go as  $\ln r /r$). The argument given there likely extends to arbitrary order, but we leave such study for the future. For the purposes of the present section however, it is enough to use the form (\ref{lamlead}). Indeed, only the leading term $\e(\xh)$ contributes to the charge. 

After recovering the known electric and magnetic-type charges for the gauge parameter (\ref{lamlead}), we review how their Ward identities correspond to Weinberg's soft photon theorem. This will serve as motivation for the analysis of  section \ref{secOr}.

\subsection{Electric-type charge}
Substituting (\ref{lamlead}) in (\ref{rholam}) and using the fall-offs (\ref{fallF}), the electric-type charge (\ref{Qlam}) at null infinity is found to be
\be
Q_\e = \int_\scri  d^3V \, \e \, \partial_u \Ftwo_{ru} \label{Qlead1}
\ee
where $d^3V = du d^2 \xh \sqrt{q}$ is the volume element on $\scri$.  We now make use of the field equations (\ref{eom}) in order to express (\ref{Qlead1})   in terms of the free data. From the leading part of the field equation $\nabla^b \F_{ub}=\j_u$ one finds:
\be
\partial_u \Ftwo_{ru} = \jtwo_u - D^A \Fo_{uA}. \label{F2ru}
\ee
On the other hand, the fall-offs discussed section \ref{secprel} imply:
\ba
j_u & = & i e \, \phi \, \partial_u \phi^* + c.c.\\
\Fo_{uA} & = & \partial_u A_A. \label{FouA}
\ea
Thus, one concludes
\be
Q_\e = \int_\scri d^3 V \, \e ( j_u - \partial_u D^B A_B),
\ee
which corresponds to the charge used in  \cite{stromqed,mohd}.
\subsection{Magnetic-type charge} \label{secmaglead}
We now consider $\lambda$ as above, with free data given by a sphere function $\et(\xh)$:
\be
\lambda(u,r,\xh) = \et(\xh) + O(r^{-\epsilon})  \label{lamet}
\ee
Substituting (\ref{lamet}) in (\ref{rholamt}) and using (\ref{fallF}) one finds the magnetic-type charge (\ref{Qlamt}) is given by:
\be
\Qt_{\et}  =   \int_{\I}  du \, d^2 \xh \, \et \, \partial_u( \eta^{AB} F_{AB})  . \label{Qtlead1} 
\ee
In terms of the free data $F_{AB}$ is simply the field strength of $A_A$, $F_{AB}=\partial_A A_B - \partial_B A_A$.
Defining 
\be
[F_{AB}](\xh) :=  F_{AB}(u=+\infty,\xh)-F_{AB}(u=-\infty,\xh) \label{brF}
\ee
the charge (\ref{Qtlead1}) may alternatively be written as:
\be
\Qt_{\et}  = \oint_{S^2} \et \, \eta^{AB}[F_{AB}]  .\label{Qet}
 \ee

\subsection{Relation to leading soft theorem} \label{seceqlead}
Repeating the steps at past null infinity, one ends up with two pair of charges $Q^\pm_\e$ and $\Qt^\pm_{\et}$ associated to future (+) and past (-) null infinity. In \cite{stromqed,mohd} it is shown that the conservation of $Q_\e$ in the S matrix sense:
\be
Q^-_\e S = S Q^+_\e \label{leadelwi}
\ee
follows from Weinberg's soft photon theorem. Conversely (\ref{leadelwi}) was shown to imply such theorem, provided certain condition on the asymptotic values of $F_{AB}$ is satisfied. In fact, the minimal condition required to go from (\ref{leadelwi}) to the soft theorem is to demand:
\be
[F^-_{AB}] = [F^+_{AB}],  \label{brFpm}
\ee
with $[F^\pm]$ as defined in (\ref{brF}) for future and past null infinity respectively. Now, looking at the expression of the magnetic charge $\Qt_{\et}$ (\ref{Qet}) it follows that  (\ref{brFpm}) is the condition for the conservation of such charge. In S matrix notation:
\be
\Qt^-_{\et} S = S \Qt^+_{\et} \label{leadmagwi}.
\ee
Finally, one can verify that condition (\ref{leadmagwi}) follows from Weinberg's soft photon theorem (see for instance discussion at the end of section 5.2 of \cite{cl3}).  To summarize: Weinberg's soft photon theorem gives two (per point on the sphere) identities associated to  the two soft photon polarizations (times each soft photon direction). These are equivalent to the two (per point on the sphere) identities (\ref{leadelwi}) and  (\ref{leadmagwi}).

\section{$O(r)$ large gauge transformations and associated charges} \label{secOr}
We now look at gauge parameter that satisfy the wave equation (\ref{boxlam}) and that diverge linearly in $r$ as one moves to null infinity. Starting with the ansatz
\be
\lambda(u,r,\xh) = r \lamone(u,\xh) + \lamzero(u,\xh) + O(r^{-\epsilon}), \label{rlam}
\ee
one finds (see appendix \ref{largeggeapp} for details)
\be
\lamone(u,\xh)  =  \mu(\xh), \quad \quad \lamzero(u,\xh)  =  u ( 1 + \Delta/2) \mu(\xh). \label{lammu}
\ee
$\mu(\xh)$ is unconstrained and plays the role of  `free data' for such large gauge transformation. $\Delta$ is the unit sphere Laplacian.

\subsection{Electric-type charge} \label{secOrel}
Substituting (\ref{rlam}) in (\ref{rholam}) and using (\ref{fallF})  one obtains 
\be
\rho_\lambda = \sqrt{q}[ \lamone \Ftwo_{ur} - r \partial_u(\lamone \Ftwo_{ur}) - \partial_u(\lamone \Fthree_{ur}+\lamzero \Ftwo_{ur})] + O(r^{-\epsilon}) , \label{rhorlam}
\ee
where we dropped the total divergence term in (\ref{rholam}). Since the limit of interest is $t \to \infty$ with $u=$constant, we set $r=t-u$ in (\ref{rhorlam}) and using (\ref{lammu}) arrive at:
\be
\rho_\lambda= t \, \rhodiv +\rhofin + O(t^{-\epsilon})
\ee
with
\ba
\rhodiv & = &  \sqrt{q} \,\mu \, \partial_u \Ftwo_{ru} \label{rhodiv}\\
\rhofin & = & \sqrt{q}[\mu \,  \partial_u \Fthree_{ru} + \frac{\Delta\mu}{2} u \partial_u  \Ftwo_{ru} + \frac{\Delta\mu}{2} \Ftwo_{ru}].  \label{rhofin}
\ea
Comparing with (\ref{Qlead1}), we see that $\rhodiv$ coincides with the charge density of a `standard' large gauge transformation $\lambda \sim \mu$. Our prescription to obtain a finite charge amounts to discard such contribution associated to the leading soft photons. We interpret this prescription as a phase space counterpart of how leading soft photons are `projected out' in Eq. (\ref{lowthm})  \cite{stromlow}.


We now focus attention in the finite charge density (\ref{rhofin}). From the leading field equations $\nabla^b \F_{a b}=\j_a$ for $a=r,A$ one finds:
\ba
\Fthree_{ru} & = & j_r + D^A \Ftwo_{A r} \label{fer} \\ 
j_A & = & - \partial_u \Ftwo_{A r} + \Fone_{A u} + D^B F_{AB} \label{feA}
\ea
Using the leading relation of the Bianchi identity $\partial_{[A} F_{r u]}=0$,
\be
\Fone_{Au}= - \partial_u \Ftwo_{Ar} - \partial_A \Ftwo_{ru} \label{bid}
\ee
in Eq. (\ref{feA}) and solving for $\partial_u \Ftwo_{A r}$ one obtains
\be
\partial_u \Ftwo_{A r}= -\frac{1}{2} j_A - \frac{1}{2} \partial_A \Ftwo_{ru} + \frac{1}{2} D^B F_{AB} \label{pufar}.
\ee
Now applying $\partial_u$ on (\ref{fer}) and using the divergence of relation (\ref{pufar}) leads to:
\be
\partial_u \Fthree_{ru} = \partial_u j_r -\frac{1}{2} D^A j_A -\frac{1}{2}\Delta \Ftwo_{ru} \label{F3ru}
\ee
where we used that $D^A D^B F_{AB}=0$ which follows from the antisymmetry of $F_{AB}$.
When (\ref{F3ru}) is used in (\ref{rhofin}) a further simplification arises: The contribution coming from the last term in (\ref{F3ru}) cancels (upon integration on the sphere) the last term in (\ref{rhofin}). Also, the $u \to \pm \infty$ fall-offs of the fields imply that  $j_r(u= \pm \infty) = 0$ and hence the first term in (\ref{F3ru}) gives a vanishing contribution to the charge (see Appendix \ref{appfallu} for details). Collecting all these results and using (\ref{F2ru}), (\ref{FouA}) one finds:
\ba
Q_{r \mu} & := & \int_\I \rhofin \label{Qrmu} \\
& = & \frac{1}{2}\int_\I d^3 V( D^A \mu  \, j_A  + u \Delta \mu( j_u - \partial_u D^B A_B) ) . \label{Qrmu2}
\ea

We now note that at leading order in perturbation theory, $j_{A}\ =\ -ie(\phi^{*}\partial_{A}\phi\ -\ \phi \,\partial_{A} \phi^{*})$ and whence when we consider the Ward identity for $Q_{r \mu}$ to leading order in perturbation theory, we will use this ``non-covariant" form of $j_{A}$. Hence from now on we will assume that the $O(e^{2})$ term in $j_{A}$ (\ref{jA}) is dropped. As we see below this  Ward identity leads to Low's soft photon theorem at tree level. It is expected that when  relating $Q_{r \mu}$ to loop corrected soft theorems, the full $j_{A}$ should be taken into account.

\subsection{Magnetic-type charge}
We now compute $\rhot_{\lambda}$ (\ref{rholamt0}) with $\lambda$ as in (\ref{rlam}) with free data $\mut(\xh)$,
\be
\lamone(u,\xh)  =  \mut(\xh), \quad \quad \lamzero(u,\xh)  =  u ( 1 + \Delta/2) \mut(\xh). \label{lammut}
\ee
From (\ref{rholamt}) and (\ref{fallF}) one finds:
\be
\rhot_{\lambda}= \eta^{AB}[ \lamone \Fo_{AB} - r \partial_u( \lamone \Fo_{AB}) - \partial_u(\lamone \Fone_{AB}+ \lamzero \Fo_{AB})] +O(r^{-\epsilon}).
\ee
As in the previous section we set $r=t-u$ and use (\ref{lammut}) to obtain
\be
\rhot_\lambda= t \, \rhotdiv +\rhotfin + O(t^{-\epsilon})
\ee
with
\ba
\rhotdiv & = & -\eta^{AB}\, \mut \, \partial_u F_{AB} \label{rhotdiv}\\
\rhotfin & = & -\eta^{AB}[\mut \,  \partial_u \Fone_{AB} + \frac{\Delta\mut}{2} u \partial_u  F_{AB} + \frac{\Delta\mut}{2} F_{AB}]  \label{rhotfin}
\ea
(recall that $F_{AB} \equiv \Fo_{AB}$). Similarly to the previous section, the divergent piece (\ref{rhotdiv}) corresponds to the `leading soft photon' magnetic charge of section \ref{secmaglead}. It now remains to express $\Fone_{AB}$ in (\ref{rhotfin}) in terms of the free data. We start with the leading relation of the Bianchi identity $\partial_{[r} \F_{AB]}=0$:
\be
\Fone_{AB} = 2 \partial_{[A} \Ftwo_{B] r} \label{bi2}.
\ee
Applying $\partial_u$ on (\ref{bi2})  and using Eq. (\ref{pufar}) one obtains
\be
\partial_u \Fone_{AB} = - \partial_{[A} j_{B]} - \frac{1}{2}\Delta F_{AB} \label{puF1AB}
\ee
where we used the identity:  $2 D_{[A} D^{C}F_{B] C} = - \Delta F_{AB}$. When using (\ref{puF1AB})  in (\ref{rhotfin}), the contribution coming from the last term in (\ref{puF1AB}) cancels (upon integration on the sphere) the last term in (\ref{rhotfin}). The final expression for the charge reads:
\ba
\Qt_{r \mut} & := & \int_\I \rhotfin  \label{Qrmut} \\
& = & \int_\I du \, d^2 \xh \, \eta^{AB}[- \mut \partial_{[A} j_{B]} + \frac{1}{2} u  \Delta \mut \partial_u F_{AB} ] .\label{Qrmut2}
\ea

\subsection{Relation to subleading soft theorem}
In \cite{stromlow}, the authors showed that Low's subleading soft photon theorem was equivalent to Ward identities of certain charges parametrized by sphere vector fields $Y^A$ which they found to be:\footnote{Our sign convention for the current $\J_\mu$ is opposite to the one in \cite{stromlow}. $\Q_Y$ here is minus $\Q_Y$ in \cite{stromlow}. }
\be
\Q_Y= \Qh_Y + \Qs_Y \label{Qstrom}
\ee
\ba
\Qh_Y & = &  \int_\I d^3 V ( u D^A Y_A j_u + Y^A j_A) \label{Qstromh} \\
\Qs_Y & = &- 2 \int_\I d^3 V (u D_z Y^z \partial_u D^{\zb}  A_{\zb} + u D_{\zb} Y^{\zb} \partial_u D^{z}  A_{z} ), \label{Qstroms}
\ea

The first step to compare both sets of charges is to decompose the sphere vector field as a sum of gradient and curl pieces:
\be
Y^A=  \frac{1}{2}D^A \mu - \epsilon^{AB} D_B \mut ,\label{Ymu}
\ee
where $\sqrt{q} \epsilon^{AB}= \eta^{AB}$ and $\epsilon^{z \zb}=i q^{z \zb}$. A straightforward computation (see Appendix \ref{appcomparison}) shows then that the charge (\ref{Qstrom}) for the vector field (\ref{Ymu}) is a sum of the electric and magnetic-type charges of the previous subsections:
\be
\Q_Y = Q_{r \mu} + \Qt_{r \mut}. \label{equalQ}
\ee
This, together with the results of \cite{stromlow} show that the (tree-level) subleading soft theorem is equivalent to Ward identities of the charges associated to the $O(r)$ large gauge transformations
\ba
Q^-_{r \mu} S & = & S Q^+_{r \mu} \\
\Qt^-_{r \mut} S & = & S \Qt^+_{r \mut} .
\ea
From this perspective, the situation is completely parallel to what happens for $O(1)$ large gauge transformations and the  leading soft theorem as discussed in section \ref{seceqlead}.

\section{Why not $O(r^{2})$ large gauge transformations?} \label{secOr2}
It is intriguing and at the same time slightly worrying that the divergent gauge parameters with asymptotic expansion
\be
\lambda(u,r,\xh) =  r \lamone(u,\xh) + \lamzero(u,\xh) + O(r^{-\epsilon}) \label{lamr}
\ee
give rise to finite charges which are conserved in the quantum theory. A natural question then arises. Why do we consider gauge parameters which only diverge linearly in $r$? What if we take an ansatz of the form, 
\be
\Lambda(u,r,\xh) = r^2 \overset{(2)}{\Lambda}(u,\xh) + r \overset{(1)}{\Lambda}(u,\xh) + \overset{(0)}{\Lambda}(u,\xh) + O(r^{-\epsilon}) \label{r^{2}-div}
\ee
which is quadratically divergent in $r$ as we approach null infinity?
Of course just as in the previous case, we expect the associated charges to be divergent on the radiative phase space. However based on our proposed prescription, it is only if the divergent terms can be associated to charges corresponding to leading or sub-leading soft photons that we can discard them (by projecting out the corresponding modes).  If the divergent terms do not admit such an interpretation, then we cannot allow such large gauge transformations. 
As we see below, this is indeed what happens in the present case and hence gauge parameters which diverge quadratically in $r$ are not allowed in our scheme.

Let us  consider the electric-type charge associated to the large gauge parameter of Eq.(\ref{r^{2}-div}).  Substituting (\ref{r^{2}-div}) in (\ref{rholam}) and using the fall-off conditions on ${\cal F}_{ab}$  one finds, after some algebra analogous to the calculations done in section \ref{secOrel}, 

\be
Q_{\Lambda} = \lim_{t\to \infty}\int_{\Sigma_t} du d^2 \xh \sqrt{q} (t^{2}\rho_{\textrm{div}}^{(2)}\ +\ t\rho_{\textrm{div}}^{(1)}\ +\ \rho_{\textrm{finite}})
\ee
with
\ba
\rhodiv^{(2)}  & = &  \partial_u (\overset{(2)}{\Lambda} \Ftwo_{ru} ) \\
\rhodiv^{(1)} & = & \partial_u(  \overset{(2)}{\Lambda} \Fthree_{ru}+ \overset{(1)}{\Lambda}\Ftwo_{ru}  -2 u \overset{(2)}{\Lambda} \Ftwo_{ru}). \label{rhodiv1}
\ea
The wave equation $\square \Lambda =0$ implies $\Lambda^{(2)}$ is independent of $u$. Let us choose it as $\Lambda^{(2)}(\hat{x})\ =\ \mu(\hat{x})$. It then follows (see Eq.(\ref{may25-1}) of
appendix \ref{largeggeapp}) that $\Lambda^{(1)}\ =\ \frac{u}{4}(\Delta + 6)\mu$ (plus an $u$-independent function on the sphere that corresponds to an $O(r)$ gauge parameter (\ref{lamr})). On substituting these functional forms in the above equations it is easy to see that although $\rho^{(2)}_{\textrm{div}}$ is the same as the electric-type charge associated to leading soft photons, no such interpretation exists for $\rho_{\textrm{div}}^{(1)}$. More in detail, (\ref{rhodiv1}) takes the form
\be
\rhodiv^{(1)} = \partial_u ( \mu \Fthree_{ru} + \frac{1}{4} u ( \Delta -2) \mu  \Ftwo_{ru}),
\ee
which clearly differs from the (finite) $\lambda \sim \mu$ electric-type charge  (\ref{rhofin}):
\be
\rhofin(\lambda \sim \mu) = \partial_u(\mu  \Fthree_{ru} + \frac{1}{2} u \Delta \mu  \Ftwo_{ru})
\ee
associated to subleading soft photons. 
Thus, according to our prescription, we are not able to discard the divergent piece and hence we cannot obtain a finite charge associated to (\ref{r^{2}-div}).  We are lead to conclude that large gauge transformations which diverge quadratically (or higher) in $r$ do not define a symmetry for massless QED.

\section{Conclusions}
Over the past few years many soft theorems have been reinterpreted as Ward identities thereby enhancing our understanding of symmetries in gauge theories and gravity. One such remarkable identification was given in \cite{stromlow},  where Low's subleading soft photon theorem was shown to be equivalent to new  symmetries of QED. The  associated charges $\Q_Y$ where found to be  parametrized by vector fields on the sphere $Y^A$.    One puzzling aspect of these charges is that 
they appear to be unrelated to large gauge transformations which have been  successful in interpreting Weinberg's soft photon theorem.  In this paper we  provided an alternative perspective 
which resolves this puzzle. In our proposal, the vector field $Y^A$ is just  a convenient way to  parametrize two functions $\mu$ and $\mut$ associated to large $O(r)$ gauge parameters $\lambda \sim r \mu$.  The charge $\Q_Y$ is then a sum of   electric and magnetic charges associated to such large  gauge parameters:
\be
\Q_Y \sim \lim_{\Sigma \to \I} \int_\Sigma d^3V \, [ \partial_a (r \mu E^a) + \partial_a (r \mut B^a)] ,\label{QYconc}
\ee
where $Y^A= D^A \mu - \epsilon^{AB} D_B \mut$. 
 In this way the leading and subleading soft photon Ward identities are put on the same footing.

Several interesting questions  remain open. Conservation of these charges in quantum theory were shown to be equivalent to Low's theorem, however to prove that classically the charges are conserved in scattering processes is an interesting and challenging task.  Another open question is whether the loop corrected version of Low's theorem can be associated to  Ward identities of  large gauge transformations.\\

The ideas presented here can also be implemented in gravity. In \cite{short} we show there is a  similar interpretation of the tree level sub-subleading soft graviton theorem as Ward identities of large diffeomorphisms.

\section{Acknowledgements}
We would like to thank Daniel Salinas for inspiring discussions. We thank Rafael Porto for helpful comments. MC is supported by Anii and Pedeciba. AL is supported by Ramanujan Fellowship of the Department of Science and Technology.

\begin{appendix}
\section{Large gauge parameters} \label{largeggeapp}
In this appendix we calculate the coefficients of the large $r$ expansion of gauge parameters. We start with an ansatz that includes the  most divergent term used in the paper:
\be
\lambda = r^2 \lamtwo + r \lamone + \lamzero + \frac{\log r }{r} \lamln +O(r^{-1}). \label{lamans}
\ee
The $\log r/r$ and $O(1/r)$ terms corresponds to  `small' gauge parameters. In the body of the paper we simply took them as $O(r^{-\epsilon})$. Indeed, a $O(r^{-\epsilon})$ fall-off is enough to guarantee a vanishing contribution to the charges. However, in order to have a  solution to the wave equation $\square \lambda=0$ we need the more specific form (\ref{lamans}). We will track the coefficients up to the order before $O(r^{-1})$. $O(r^{-1})$ parameters behave like regular scalar fields that satisfy the wave equation (for instance those admitting Fourier expansion) and are associated to small gauge parameters that have their own `free data'. As we will see, the  $\log r /r$  is needed for a consistent solution to the wave equation. 

Applying the wave operator to (\ref{lamans}) one gets:
\begin{multline}
\square \lambda = r[- 6 \partial_u \lamtwo]  + [(\Delta+6) \lamtwo - 4 \partial_u \lamone] + r^{-1}[(\Delta+2) \lamone -2 \partial_u \lamzero] + \\ r^{-2}[ \Delta \lamzero - 2 \partial_u \lamln] +O(r^{-3} \log r)
\end{multline}
 (for the computation it is convenient to write the wave operator as  $\square \lambda = r^{-1}(\partial^2_r - 2 \partial_r \partial_u +r^{-2} \Delta) (r \lambda)$). The general solution to $\square \lambda =0$ at the order we are working is then:
\ba\label{may25-1}
\lamtwo(u,\xh) & = & \nu(\xh) \\
\lamone(u,\xh) & = & \frac{1}{4}\int_{0}^u du' ((\Delta+6) \lamtwo) + \mu(\xh) \\
\lamzero(u,\xh) & = & \frac{1}{2} \int_{0}^u du' ((\Delta+2) \lamone) + \e(\xh) \\
\lamln(u,\xh) & = & \frac{1}{2} \int_{0}^u du' (\Delta \lamzero) + \eta(\xh) \\
\ea
At each order there appears an integration `constant' that is a function on the sphere. The $O(1)$ large gauge parameter of section \ref{secO1} correspond to $\nu=\mu=0$. The $O(r)$ parameter of section \ref{secOr} corresponds to $\nu=\e=0$ (one could have kept $\e \neq 0$ but that just gives a $O(1)$ gauge parameter). The value of $\eta$ is pure gauge. The $\log r /r$ is however crucial for otherwise one would have gotten $\Delta \lamzero=0$ which would have eliminated the $O(1)$ large gauge transformation.

\section{$u \to \infty$ fall-offs} \label{appfallu}
In this section we explicit the assumed $u \to \pm \infty$ fall-offs underlaying the analysis of section \ref{secOr}. First, in order for the `soft' charges to be well defined, we assume that 
\be
A_A=O(u^{-1-\epsilon})
\ee
For the `hard' charges, the term with slower fall-off is the one proportional to $u \phi \partial_A \phi^*$. In order for its integral to be well defined we assume
\be
\phi=O(u^{-1-\epsilon}). \label{fallphi}
\ee
With these fall-offs, all expressions of charges in section \ref{secOr} are well defined. We finally show that they also imply $j_r=O(u^{-1-\epsilon})$, which implies there is no contribution from $j_r$ in the charge (\ref{Qrmu2}). To study the fall-offs of $j_r$ (\ref{jr}) we need to express it in terms of the free data. Looking at the $O(r^{-3})$ coefficient of scalar field equation (\ref{eom2}) and using the Lorenz gauge condition one obtains the following expression for $\phitwo$:
\be
\phitwo = \phitwo_{1} - i \Atwo_r \, \phi \label{phi2}
\ee
where 
\be
 \phitwo_1 = - \frac{1}{2} \int^u_{-\infty} du'( \Delta \phi - 2 i D^B(\phi A_B)- A^B A_B \phi) +f(\xh)
\ee
with $f(\xh)$ an integration `constant'. Substituting (\ref{phi2}) in (\ref{jr}) we see that the $\A_r$ term cancels out and arrive at
\be
j_r  =  i e  \, \phi^* \phitwo_1 + c.c. \label{jr2}
\ee
Since at most  $\phitwo_1=O(1)$ we conclude that $j_r$ has the same fall-offs as $\phi$ (\ref{fallphi}.)

\section{Comparison between $Y^A$ and large $O(r)$ charges} \label{appcomparison}
We first show that:
\be
\Q_{\frac{1}{2}D^A \mu} = Q_{r \mu}. \label{compgrad}
\ee
Setting 
\be
Y^A= \frac{1}{2} D^A \mu \label{Ygradmu}
\ee
in (\ref{Qstromh}) one immediately recovers the `hard' part of (\ref{Qrmu2}) (the terms proportional to currents). To compare the soft parts, we note that for $Y^A$ as in (\ref{Ygradmu}) we have:
\be
D_z Y^z = D_{\zb} Y^{\zb} = \frac{1}{4}\Delta  \mu \label{DzYz}
\ee
since 
\be
D_z D^z \mu = D_{\zb} D^{\zb} \mu = \Delta \mu/2. \label{lapmu}
\ee 
Using (\ref{DzYz}) in (\ref{Qstroms}) one recovers the `soft' part of (\ref{Qrmu2}) (the terms proportional to $A_A$). This establishes the equality (\ref{compgrad}). We now show that 
\be
\Q_{\epsilon^{BA} \partial_B \mut} = \Qt_{r \mut}. \label{compcurl}
\ee
For
\be
Y^A= \epsilon^{BA}\partial_B \mut \label{Ycurl}
\ee
we have that $D_A Y^A=0$ and so
\be
\Qh_{\epsilon^{BA} \partial_B \mut} =\int_\I d^3 V ( \epsilon^{B A} \partial_B \mut  j_A). \label{Qhcurl}
\ee
Noting that $d^3 V= du d^2 \xh \sqrt{q}$ and $\sqrt{q} \epsilon^{BA}= \eta^{BA}$  and doing an integration by parts one sees that (\ref{Qhcurl}) coincides with the hard part of (\ref{Qrmut2}). We now compare the soft parts. Using that $\epsilon^{z \zb}=i q^{z \zb}$, the soft part of (\ref{Qrmut2}) can be written as:
\be
\Qt^{\text{soft}}_{r \mut}= i \int_{\I} d^3 V \, u \, \Delta \mut \, \partial_u(D^{\zb} A_{\zb}- D^z A_z). \label{Qtsoft}
\ee
Next, we note that for $Y^A$ as in (\ref{Ycurl}) we have
\be
\Delta \mut = 2 i D_z Y^z = -2 i D_{\zb} Y^{\zb} \label{lapmut}
\ee
which follows from Eq. (\ref{lapmu}) and $Y^{z}=-i D^z \mut$. Using (\ref{lapmut}) in (\ref{Qtsoft}) one verifies that (\ref{Qtsoft}) coincides with the soft charge (\ref{Qstroms}). This establishes the equality (\ref{compcurl}). Combined with (\ref{compgrad}) it gives Eq. (\ref{equalQ}).

\end{appendix}

\end{document}